# Design of Complex Experiments Using Mixed Integer Linear Programming

Storm Slivkoff[a] and Jack L. Gallant[a,b,c,+]

[a] Department of Bioengineering, University of California, Berkeley, CA 94720, USA
[b] Department of Psychology, University of California, Berkeley, CA 94720, USA
[c] Helen Wills Neuroscience Institute, University of California, Berkeley, CA 94720, USA
[+] Corresponding Author. Email: gallant@berkeley.edu

S. Slivkoff's work on this project was supported by ONR MURI N000141410671

December 3, 2020

## Abstract

Over the past few decades, neuroscience experiments have become increasingly complex and naturalistic. Experimental design has in turn become more challenging, as experiments must conform to an ever-increasing diversity of design constraints. In this article we demonstrate how this design process can be greatly assisted using an optimization tool known as Mixed Integer Linear Programming (MILP). MILP provides a rich framework for incorporating many types of real-world design constraints into a neuroimaging experiment. We introduce the mathematical foundations of MILP, compare MILP to other experimental design techniques, and provide four case studies of how MILP can be used to solve complex experimental design challenges.

## 1. Introduction

Many different tools have been used for designing neuroscience experiments, including combinatorial designs (Friston et al. 1997; Friston et al. 1998; Henson 2007), optimal designs (Dale 1999; Wager and Nichols 2003), and m-sequences (Sutter 1987; Benardete and Victor 1994; Buračas and Boynton 2002). These tools work exceedingly well for their intended use cases, but they are unable to meet many of the design challenges that arise as neuroimaging experiments become increasingly complex and naturalistic. In particular, complex neuroimaging experiments must conform to many idiosyncratic design constraints that reflect the structure of the real world. Conventional neuroimaging experimental design tools are usually specialized to accommodate a very limited set of design constraints, and they have little flexibility for integrating additional constraints into an experimental design.

A more flexible approach is to explicitly frame experimental design as a constrained optimization problem. Given a set of design constraints, a researcher wants to create an experiment that optimally tests some scientific hypotheses. This is the framing used by Mixed Integer Linear Programming (MILP). MILP is an optimization technique that is well-suited to representing the structure of complex experiments. MILP has already been used to design experiments in fields such as hydrology (Hsu and Yeh 1989), genetics (McClosky and Tanksley 2013), agricultural economics (Chou 1961), and educational assessment (Linden et al. 2004). MILP has also been used in the context of neuroimaging to phase unwrap fMRI data (Jenkinson 2003), to create a sampling prescription for diffusion imaging (Cheng et al. 2014), to analyze fMRI resting state data (Demertzi et al. 2014; Costa et al. 2015; Henry and Gates 2017; Costa et al. 2019), and to perform source localization in MEG data (Freschi 2010). However, we have been unable to identify any use of MILP in the design of functional neuroimaging experiments.

This article develops and demonstrates a conceptual framework for using MILP to design neuroimaging experiments, but the information provided here can also be applied to other types of neuroscience experiments. **Section 2** describes some of the challenges associated with the design of complex neuroimaging experiments. **Section 3** introduces the mathematical foundations of MILP. **Section 4** explores four neuroimaging case studies where MILP is uniquely equipped to optimize experimental designs. **Section 5** compares MILP to other experimental design techniques and discusses some of its limitations.

## 2. Challenges of Designing Complex Experiments

Although it is often desirable for scientific experiments to be as simple as possible, many functions of the brain can only be investigated using complex experiments. This is because the brain often responds in a qualitatively different manner when the stimuli or tasks used in an experiment are overly-simplified or unnatural (Wu and Gallant 2006). Oversimplification of an experiment can undermine its ecological validity and undermine its generalizability to real-world situations (Schmuckler 2001; Dan and Poo 2004; Wu and Gallant 2006; Sonkusare et al. 2019). Qualitative differences in brain responses evoked by simple vs complex experiments have been demonstrated in many brain subsystems, including vision (Dan and Poo 2004; David 2004), audition (Rieke et al. 1995; Woolley et al. 2005; Wild et al. 2017), language comprehension (Vandenberghe et al. 2002; Huth et al. 2016), face perception (Allison et al. 2000; Kilgour et al. 2005; Gauthier et al. 2009), social cognition (Zaki and Ochsner 2009; Hari et al. 2015; Babiloni and Astolfi 2014), and memory (Nielson et al. 2015; Ren et al. 2018; Kauttonen et al. 2018). Detailed discussions of when simple or complex experiments are most appropriate can be found in (Schmuckler 2001; Dan and Poo 2004; Sonkusare et al. 2019).

To avoid problems related to oversimplification, complex experiments must incorporate the rich structure of the real world into their designs. However, conventional tools for designing neuroimaging experiments are often not expressive enough to represent this structure. These conventional tools include m-sequences (Sutter 1987; Benardete and Victor 1994; Buračas and Boynton 2002), optimal designs (Dale 1999; Wager and Nichols 2003), and combinatorial designs, where combinatorial designs are a broad category that includes factorial designs (Friston et al. 1997; Hall et al. 2000; Gurd et al. 2002), latin squares (Friston et al. 1998; Hagberg et al. 2001), and many others. Each of these tools can be useful for selecting and arranging the stimuli, tasks, and other types of conditions in an experiment. Each tool can imbue an experiment with desirable properties such as efficiency and precision. However, when experiments grow in complexity, these tools can become difficult or impossible to use. In this section we will describe some of the challenges associated with designing complex experiments.

### 2.1. High Dimensionality

One common motif of complex experiments is that they often have a large number of variables in their designs. Thus, designing a complex experiment often requires many decisions about the selection and arrangement of experimental stimuli, tasks, and other experimental variables. For example, it is common for vision studies to utilize hundreds or thousands of stimuli (Thorpe et al. 1996; Kay et al. 2008). Cognitive control experiments might require subjects to learn tens or hundreds of novel mappings between stimuli and task-relevant variables (Gauthier et al. 1998; Badre et al. 2010; Dresler et al. 2017). In each of these cases, the researcher is required to make many decisions about how to select and arrange the tasks and stimuli in their experiment.

Each design decision might also depend on a large number of variables. For example, natural speech or natural movie stimuli might contain hundreds or thousands of unique features that affect the resultant brain activity (Nishimoto et al. 2011; Huth et al. 2012; Sudre et al. 2012; Huth et al. 2016; Heer et al. 2017). Optimally selecting or arranging stimuli can thus require accounting for a large number of such features. Even a synthetic dot-motion experiment can have many relevant experimental parameters, including color, color coherence, motion direction, motion speed, motion coherence, dot count, dot size, cluster size, or fixation position (Newsome and Pare 1988; Mante et al. 2013).

Conventional neuroimaging design tools do not adequately scale to the large number of dimensions associated with complex experimental designs. Even in simple low-dimensional cases, tools such as m-sequences or combinatorial designs can impose undesirable constraints on the length of the experiment or the way in which experimental conditions must combine (Vieira et al. 2011). These difficulties only grow worse as an experiment grows in complexity (Vieira et al. 2011). Tools such as fractional factorial designs and balanced incomplete block designs have been developed to provide better scalability in the face of high dimensionality (Montgomery 2008). However, these can still only scale to relatively small numbers of dimensions. They also suffer from the other experimental design challenges that we describe below. An ideal experimental design tool would flexibly scale to whatever number of dimensions are relevant to the experiment.

## 2.2. Idiosyncratic Structure

Another challenge is that each individual dimension of a complex experiment might have its own idiosyncratic structure. In other words, each of the experimental conditions (related to the stimuli, tasks, or other variables) might have unique characteristics that require special treatment by the experimental design. For example, consider an experiment comprised of simple binary contrasts, such as those used to localize cortical regions of interest like the Fusiform Face Area or the Parahippocampal Place Area (Saxe et al. 2010). In this type of experiment each experimental condition has a simple structure. Each condition has two possible levels, on or off. Each condition is able to precede or follow each other condition. Each condition can be treated identically in the experimental design.

Now compare this to the design of a complex executive control experiment where the subject navigates a decision tree. Decision trees have been used in studies of gambling (Paliwal et al. 2014), foraging (Kolling et al. 2012), and in other executive control tasks (Daw et al. 2005; Gläscher et al. 2010). Here the subject must make a series of decisions, where each decision has specific consequences for the experimental states that follow. Unlike the binary contrast experiment, each of these decision conditions can have complex and idiosyncratic structure. Each decision might have its own unique set of other decisions that it can precede or follow. Each might also have its own unique effects on subsequent stimuli, rewards, or other experimental parameters.

Another example of complex idiosyncratic structure can be found in navigation experiments. Consider a navigation experiment where subjects are instructed to navigate to different locations on a map (Hartley et al. 2003) (Spiers and Maguire 2006). Depending on the hypotheses being tested, it might be important for the experimental design to account for the unique structure of the map, including its particular destinations, landmarks, and routes. The structure of a map is most naturally represented as spatial data or graph data. However, neither spatial data nor graph data can be directly represented using conventional neuroimaging design tools such as m-sequences or combinatorial designs. Failing to account for the full structure of the experimental paradigm could lead to suboptimal designs.

Conventional experimental design tools excel when each dimension of the experiment has simple structure and can be treated identically. For binary contrast experiments, it would be straightforward to use tools such as m-sequences or factorial designs to designate how the experimental conditions should be ordered and combined. These tools can fully represent the relevant structure of the binary contrast experimental paradigm. However, these tools can struggle to represent decision tree experiments, navigation experiments, or any other experiment that has complex idiosyncratic structure. An ideal experimental design tool would be able to directly represent and manipulate whatever structures are relevant to the experiment.

## 2.3. Incorporation of Pre-existing Media

There are also challenges associated with the common practice of incorporating pre-existing media into neuroimaging experiments. This has been done with movies (Nishimoto et al. 2011), images (Thorpe et al. 1996), books (Wehbe et al. 2014), music (Khorram-Sefat et al. 1997), podcasts (Huth et al. 2016), maps (Nielson et al. 2015), and video games (Mathiak and Weber 2006). Using pre-existing media can enable many novel types of experiments. However, because these media were originally created for a different purpose, they often have undesirable limitations that dictate how they can be used. For example, the media might contain a limited number of exemplars of whatever phenomena are being studied, or it might contain undesirable confounds that prevent certain hypotheses from being tested. The available source materials can impose significant constraints on an experiment, and these constraints must be considered throughout the experimental design process.

Most experimental design tools are unable to directly represent limitations in the available source materials. Instead, the researcher must separately verify that the designs produced by these tools are compatible with whatever source materials are available. Experimental design often becomes an iterative process of adjusting design parameters and checking whether the resultant designs are both feasible and optimal for the given source materials. For researchers in the exploratory phase of designing their experiments, this iterative process can make it difficult to assess the full range of experiments that are possible given the available source materials. Searching through possible designs can be even more challenging when experiments suffer from the high dimensionality or idiosyncratic structure described in previous sections. An ideal experimental design tool would directly represent the available source materials so that those materials can be used optimally.

## 2.4. Diverse Interdependent Constraints

Finally, it is important to mention that design constraints for a single experiment can be highly diverse and interdependent. We have already mentioned possible design constraints relating to high-dimensionality, idiosyncratic structure, and pre-existing media, but many other factors can constrain the design of an experiment. For example, almost all experiments have restrictions on the total amount of data that can be collected due to subject endurance, patient availability, and research budgets. The specific hypotheses being tested can also constrain experimental designs. For example, a study of high-level visual semantics might need to control for low-level features such as luminance to help ensure that the observed effects are caused by semantic processing (Dell'Acqua and Grainger 1999). Many neuroimaging experiments also seek to account for known properties of the system, such as hemodynamics and spatiotemporal resolution of the imaging modality (Amaro and Barker 2006).

Integrating diverse constraints into a single experimental design can be challenging because the constraints often interact or conflict with one another. For example, constraining the experimental timing parameters to account for hemodynamics might affect the total number of trials that can be collected. Limiting the total number of trials might limit the number of possible experimental conditions, because each condition might require some minimal number of trials. Because each condition might have its own media requirements, the possible experimental conditions are also constrained by whatever pre-existing media are available. Pre-existing media with a temporal component such as videos or audio can also place constraints on the experimental timing parameters. The interdepence of these design constraints makes them difficult to address independently.

Conventional design tools are not well-suited for integrating many diverse constraints into a single design. Instead, each of these tools tends to be specialized for optimizing one specific type of constraint. For example, m-sequences are useful for determining how conditions should be optimally ordered in an experiment. Factorial designs are useful for determining how conditions should be optimally combined. Imposing additional types of constraints on these designs requires extra steps. One approach is to generate many candidate designs that obey a subset of the design constraints, and then pick whichever candidate best satisfies the remaining constraints. This method has been used in (Kao et al. 2009) to produce a design that most efficiently accounts for the known properties of the BOLD signal. Another approach is to directly adjust the available free parameters in a conventional design to best approximate the remaining constraints. Both of these approaches are suboptimal because they prioritize some constraints over others and they do not make tradeoffs between different constraints explicit. They also optimize each component of the experiment separately, missing out on possible synergies that can come from jointly optimizing all of the components together (Glasmachers 2017). An ideal experimental design tool would represent all of its constraints within a single framework so that any tradeoffs are made explicit and all aspects of the problem can be optimized jointly.

## 3. Mathematical Foundations of MILP

Mixed Integer Linear Programming (MILP) is an optimization technique that is well-suited to the complex experimental design challenges described in the previous section. **Figure 1** shows an overview of how MILP is used in this context. On a broad level, the researcher expresses their experimental design problem as a mixed integer linear program, which a MILP solver then uses to search for an optimal experiment.

The mathematical formulation of MILP is as follows. Each mixed integer linear program specifies a vector of variables $x$ (**Figure 1.i**). In the context of experimental design, these variables are used to track various quantities related to the design. For example, there might be a variable that tracks the value of each timing parameter, or there might be a variable that tracks how many times each experimental condition is measured. For each variable in $x$, the program specifies whether the variable is a real number or an integer.

Each program also has linear constraints that specify which values the $x$ variables are allowed to take (**Figure 1.ii**). These constraints are in the form of linear equalities as $Ax = b$ or linear inequalities as $Ax \leq b$ (as shown later in this section, these two representations are interchangeable). These constraints are used to represent various design specifications of the experiment. For example, a researcher might need to specify the valid range of values for each timing parameter, or they might need to place a lower bound on how many times each experimental condition should be measured. Every vector $x$ that satisfies the integrality constraints and all of the linear constraints specified by $A$ and $b$ is considered a valid experimental design. The set of all such $x$ vectors is known as the feasible set.

The final component of each program is a linear cost function $c^T x$ that is used to evaluate the quality of each feasible $x$ vector (**Figure 1.iii**). For our purposes, the cost function is used to select an experimental design that best achieves some quality of interest. For example, a researcher might want to minimize the value of some timing parameter. Or they might want a design that best balances the number of times that each condition is sampled. In each case a cost function is used to prioritize the quality of interest. A vector $x^*$ that minimizes $c^T x$ while satisfying all of the constraints is an optimal solution to the program. In the context of experimental design, $x^*$ consists of parameter values that create an optimal experiment.

MILP is well-suited to the challenges described in the previous section: For large problems, MILP can scale to many thousands of variables and constraints. For problems where each condition has its own unique requirements, MILP can set unique constraints for each variable. For problems that depend on a pre-existing media dataset, MILP can create unique variables and constraints for each item in that dataset. For problems that have diverse design specifications, MILP can express and combine many different types of constraints. For problems with design specifications that are overly-rigid or contradictory, MILP can approximate constraints and set the relative importance of each constraint individually.

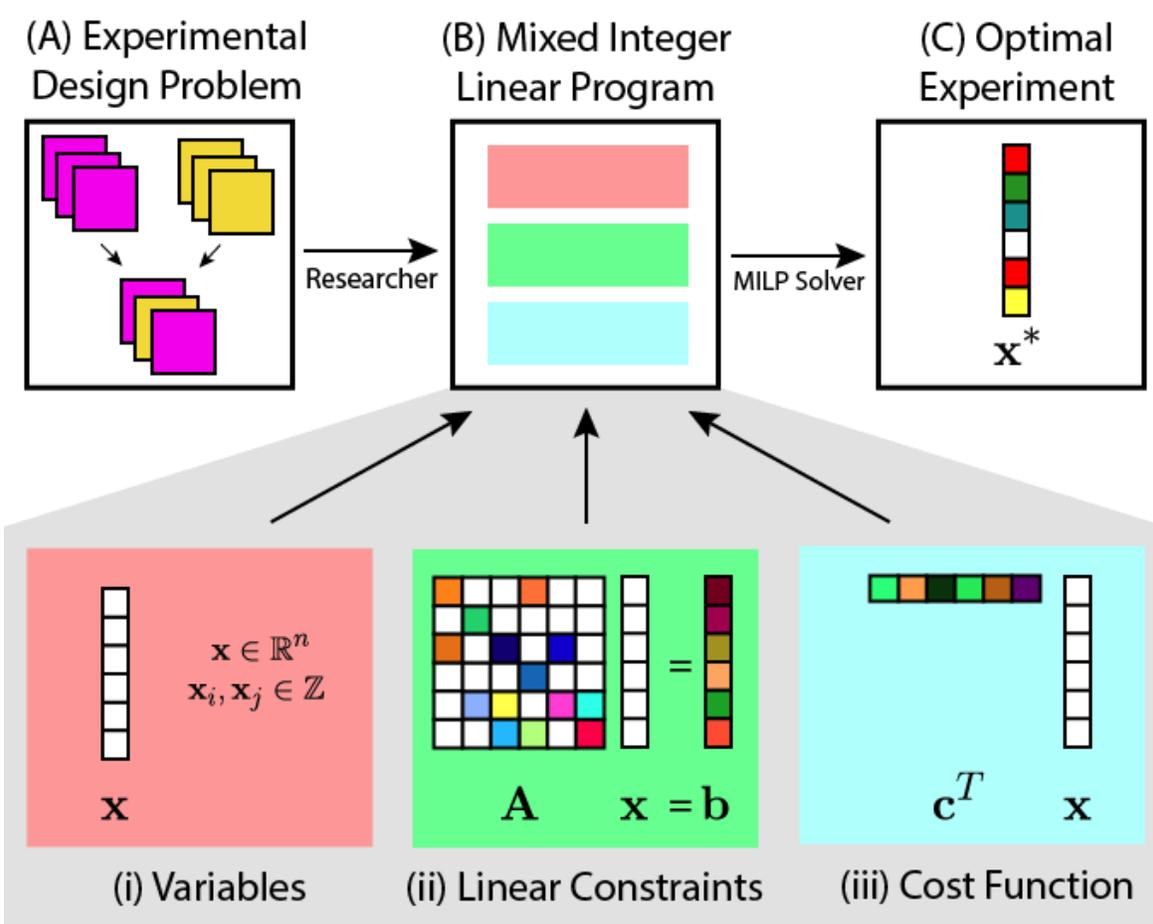

**Figure 1: Use of MILP for Experimental Design.**
**(A)** MILP is an optimization technique that is applicable to many experimental design problems. MILP can represent many different types of design constraints and incorporate them into an optimal experimental design. **(B)** To use MILP for experimental design, the researcher must first express the structure of their design problem as a mixed integer linear program. This program will consist of variables, linear constraints, and a cost function. **(i)** Variables are used to track various quantities and decisions related to the design of the experiment. Each variable encodes some aspect of the design as either a real number or an integer. For example, an integer variable might be used to track how many times an experimental condition is sampled. All of the variables are arranged into a vector $x$. **(ii)** Linear constraints specify the values that the variables are allowed to take. For example, these constraints could specify the minimum number of trials required for each condition. Each value of $x$ that satisfies all of these constraints is a valid experimental design. **(iii)** Finally, a cost function evaluates the overall desirability of each possible design. For example, the cost function might place high value on designs that evenly balance the number of trials for each condition. **(C)** Finally, a MILP solver software package is used to find an optimal solution to the program. This solution consists of values for each of the variables in $x$. These values specify an experimental design that satisfies all given constraints in a globally optimal way.

## 3.1. Solving Mixed Integer Linear Programs

Although mixed integer linear programs can be difficult to solve, there are many efficient solution algorithms and software packages available (Cook 2010; Bixby 2012). Most state of the art solvers are commercial and proprietary, but free licenses are usually available to the academic community. The examples in this article were solved using GUROBI (Gurobi Optimization 2015).

The two most common algorithms for solving mixed integer linear problems are cutting planes and branch-and-bound (Bradley et al. 1977). The cutting plane method adds auxiliary constraints iteratively until non-integral relaxations of the problem provide a solution. The branch-and-bound method divides the problem iteratively into subproblems that are solved in parallel. These two approaches are often combined into a branch-and-cut method, and they are also used alongside a variety of presolving, preconditioning, and other heuristics (Cook 2010; Bixby 2012). Details of these techniques are beyond the scope of this article, but further information is widely available (Bradley et al. 1977; Newman and Weiss 2013).

After an end user of MILP expresses their problem as a mixed integer linear program, a solver can take that program and search for a solution. Although some solvers provide tuning parameters, in many cases the end user does not need to provide any information about how the program should be solved. Given enough time, the solver will find a solution unless the problem is numerically unstable, intractable, or infeasible (see **Section 5** for details about when these conditions can occur). Importantly, once a MILP program is solved, the solution will be a global optimum. This is in contrast to some other fields of optimization where solutions might be obtained with no guarantee of global optimality.

## 3.2. Expressing Experimental Design Problems as Programs

To solve an experimental design problem using MILP, it is necessary to express each aspect of the problem using some combination of variables, linear constraints, and a linear cost function. The resulting program will then consist of all of the variables to form the vector *x*, all of the linear constraints to form the matrix *A* and vector *b*, and all of the cost function terms to form the vector *c* (**Figure 1**).

Before giving concrete examples of real-world programs, it is necessary to introduce some of the common techniques for representing problems using MILP. Many such techniques exist, but the remainder of this section focuses on the ones that are particularly useful for experimental design. Using MILP in this context then becomes a matter of recognizing how to express each experimental design specification using these techniques. Although each of these techniques may appear to be mathematically abstract, we will demonstrate their concrete usage in the real-world case studies featured in the next section.

*Interchanging equalities with inequalities*

Linear constraints can either take the form of equalities (*Ax* = *b*) or inequalities (*Ax* ≤ *b*). Many MILP solvers allow for both types of constraints, but for those that do not, explicit conversions can be used. Any equality constraint can be expressed using two inequality constraints:

> CONSTRAINTS
> $$Ax = b \leftrightarrow (Ax \leq b \text{ and } Ax \geq b)$$

Similarly, inequality constraints can be expressed using equality constraints by introducing non-negative slack variables *s*:

> CONSTRAINTS
> $$Ax \leq b \leftrightarrow Ax + s = b \text{ and } s \geq 0$$
> and

$$Ax \geq b \leftrightarrow Ax - s = b \text{ and } s \geq 0$$

It is also possible to interchange greater-than constraints with less-than constraints using simple negations if that is what a real-world problem calls for:

> CONSTRAINTS
> $$Ax \geq b \leftrightarrow (-A)x \leq (-b)$$

*Minimizing maxima*

The standard MILP cost function takes the form

> COST FUNCTION
> $$\min c^T x$$

Suppose it is desirable to include a maximum term in the cost function:

> COST FUNCTION
> $$\min c^T x + \max(y_1^T x + z_1, y_2^T x + z_2, y_3^T x + z_3))$$
> where $y_1, y_2, y_3$ are vectors and $z_1, z_2, z_3$ are scalars

This can be accomplished by introducing an auxiliary variable $v$ with constraints

> CONSTRAINTS
> $$v \in \mathbb{R}$$
> $$v \geq y_1^T x + z_1$$
> $$v \geq y_2^T x + z_2$$
> $$v \geq y_3^T x + z_3$$

and then placing $v$ in the cost function as

> COST FUNCTION
> $$\min c^T x + v$$

This process can be repeated to include multiple maxima terms in the cost function. Minima terms can be included in the cost function in certain circumstances using techniques known as big-M methods (Camm et al. 1990), however, these are beyond the scope of this article.

*Constraining maxima*

Suppose it is desirable to constrain an expression to be greater than the maximum of other expressions:

> CONSTRAINTS
> $$\max(y_1^T x + z_1, y_2^T x + z_2, y_3^T x + z_3) \leq y_4^T x$$
> where $y_1, y_2, y_3, y_4$ are vectors and $z_1, z_2, z_3$ are scalars

This can be accomplished by introducing a set of inequalities

> CONSTRAINTS

$$y_1^T x + z_1 \leq y_4^T x$$
$$y_2^T x + z_2 \leq y_4^T x$$
$$y_3^T x + z_3 \leq y_4^T x$$

A similar process can be used to constrain an expression to be less than the minima of other expressions. Constraining the opposite direction (i.e. less than a max or greater than a min) is again possible using big-M methods (Camm et al. 1990).

*Minimizing or constraining absolute values*

Absolute values can be expressed using maximization or minimization:

$$|y^T x + z| = \max(y^T x + z, -(y^T x + z))$$
$$= -\min(y^T x + z, -(y^T x + z))$$

Thus, absolute values can be used in constraints or in cost function terms using the previously described techniques

COST FUNCTION
$$\min |y^T x + z| \leftrightarrow \min \max(y^T x + z, -(y^T x + z))$$

CONSTRAINTS
$$|y_1^T x + z_1| \leq y_2^T x + z_2$$
$$\leftrightarrow \max(y^T x + z_1, -(y^T x + z_1)) \leq y_2^T x + z_2$$

This can be used to push a variable $v$ toward a target value $z$:

COST FUNCTION
$$\min |v - z|$$

It can also be used to constrain a variable v to fall within some distance $d$ of a target value $z$:

CONSTRAINTS
$$|v - z| \leq d$$

*Balancing quantities by minimizing absolute deviation*

Suppose a problem requires that a set of variables $\{v_1, v_2, ..., v_n\}$ assume similar values. There are many ways this can be accomplished. One way is to minimize the absolute deviation of each variable from the overall mean. Let µ equal the mean of these variables. Then the cost function is merely the sum of absolute deviations from this mean

COST FUNCTION
$$\mu = \frac{1}{n} \sum_{i=1}^{N} v_i$$

$$\min \sum_{i=1}^{n} |v_i - \mu|$$

More generally, this technique can be used to balance linear expressions rather than just balancing individual variables.

*Balancing quantities by minimizing range*

Another way to balance the variables $\{v_1, v_2, ..., v_n\}$ is to minimize their range, i.e.

<u>COST FUNCTION</u>
$$\min \left\{ \max_i(v_i) - \min_i(v_i) \right\}$$

These extrema can be included in the cost function using the previously described techniques. Here it is possible to include a minimization term in the objective function because it has a negative coefficient, and thus it behaves similarly to a maximization term. Using the range to balance a set of variables is useful in cases where it is desirable to avoid outliers. Multiple types of balance can be used simultaneously depending on the needs of the problem.

*Binary variables*

A variable that is constrained to be integral and fall in the range [0, 1] is known as a binary variable. In some contexts this is also known as an indicator variable or or dummy variable. Mixed integer linear programs can use binary variables to represent binary decisions. The causes or effects of those decisions can be controlled using weighted sums of binary variables in constraints or cost function terms. For example, if a program needs to place items into groups, a binary variable could be introduced for each possible {item, group} pair. Weighted sums of these variables can then be used to constrain or optimize group properties. Another example is when a program must arrange items into an optimal order. In this case a binary variable can be introduced for each possible {item, position} pair.

## 4. Experimental Design Case Studies

This section provides four case studies that illustrate how MILP can be used to solve realistic challenges in the design of complex experiments. We focus here on neuroimaging experiments, but the general principles are applicable to other types of experiments as well. These examples are intended to demonstrate the power and flexibility of MILP. Each can be solved using the basic techniques that were introduced in the previous section. The full code used to generate each example is available at https://github.com/gallantlab/milp_experimental_design .

### 4.1. Case Study 1: Balanced Grouping

A common challenge when designing experiments is to distribute conditions across blocks or sessions in a balanced way. Perhaps a researcher requires that each block of the experiment must have a similar number of exemplars of some stimulus category. Perhaps a researcher must collect data on a large battery of tasks, and the task distribution should be similar across each scanning session. Perhaps a researcher needs to evenly distribute content across trials, blocks, and sessions simultaneously. All of these scenarios can be addressed by MILP in a similar manner.

*Design Specifications*

In this first example we will cover the simple case of trying to balance the mean luminance of visual stimuli across scanning runs. Suppose a researcher is designing a vision study using 360 short videos. Each video is 20 seconds long and has its own mean luminance. For data collection, the videos must be grouped into 12 separate runs of 10 minutes each, with each video appearing in exactly one run. Importantly, the researcher plans to perform an analysis that requires all of the runs to have the same mean luminance. Although this may seem like a trivial requirement, it can be extremely challenging in practice. There are $10^{367}$ possible groupings that split 360 items into 12 equally sized groups. This space is far too large to test every possible grouping. The solution space is also discrete, so gradient-based methods cannot be used. **Figure 2B** shows a synthetic distribution of luminances that we have created for the videos.

*MILP Formulation*

It is straightforward to express this problem as a mixed integer linear program. For notation, let $V$ be the number of videos, $R$ be the number of runs, $v$ be the index over videos, and $r$ be the index over runs. The first step is to define the variables of the solution space. Introduce binary variables $X$ to represent possible pairings between runs and videos:

<u>VARIABLES</u>
$$X \in \mathbb{B}^{V \times R}$$
$$x_{v,r} = 1 \rightarrow \text{ video } v \text{ is in run } r$$
$$\phantom{x_{v,r}} = 0 \rightarrow \text{ video } v \text{ is not in run } r$$

All of the desired grouping properties in this problem can be expressed as linear equalities over elements of $X$. To specify that each run must contain $V / R = 30$ videos, introduce a constraint for each run:

<u>CONSTRAINTS</u>
$$\sum_v x_{v,r} = 30 \quad \forall \ r \in \{1, ..., R\}$$

To specify that each video appears in exactly one run, introduce a constraint for each video:

<u>CONSTRAINTS</u>
$$\sum_r x_{v,r} = 1 \quad \forall \ v \in \{1, ..., V\}$$

Any value of $X$ that satisfies all of the above constraints is a feasible grouping that produces a valid experiment.

A cost function can now be designed to select the experiment that best balances luminance. Assume that the mean luminance of each video is stored in a vector $f$. This vector is predetermined by the video dataset. Use a cost function that is the sum of absolute deviations between the global mean and the mean of each run.

<u>COST FUNCTION</u>
$$\mu = \frac{1}{V} \sum_v f_v$$
$$= \text{mean luminance across all videos}$$
$$\mu_r = \frac{R}{V} \sum_v f_v \, x_{v,r}$$
$$= \text{mean luminance within run } r$$

$$\min \sum_r |\mu - \mu_r|$$

**Figure 2C** and **Figure 2D** show a solution to this program. As guaranteed by the MILP solver, this solution achieves the lowest cost function value of any $X$ in the feasible set. **Figure 2C** shows the degree to which the mean luminance of each run deviates from the global mean luminance. This value is small compared to the overall range of luminances. **Figure 2D** compares the solution found using MILP to $10^8$ solutions found using randomization. Each randomization solution is generated by simply shuffling the videos and then splitting into 12 sequential groups. The graph shows that the solution found using MILP is substantially more balanced than any solution that can be found by randomization. The best randomized solution is the one with the lowest total balance error. As shown in in Figure 2D, the best randomized solution still has a total balance error that is 116 times larger than the solution found using MILP.

**(A)** Balanced Grouping 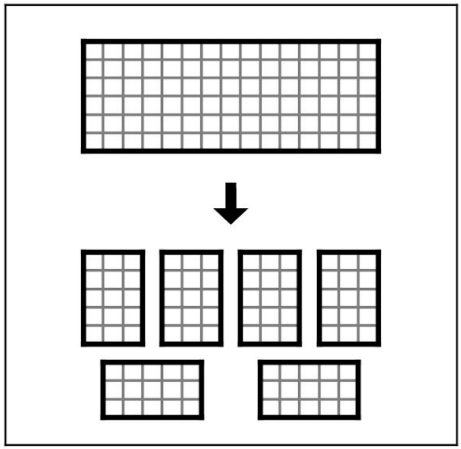

**(B)** Luminance Distribution 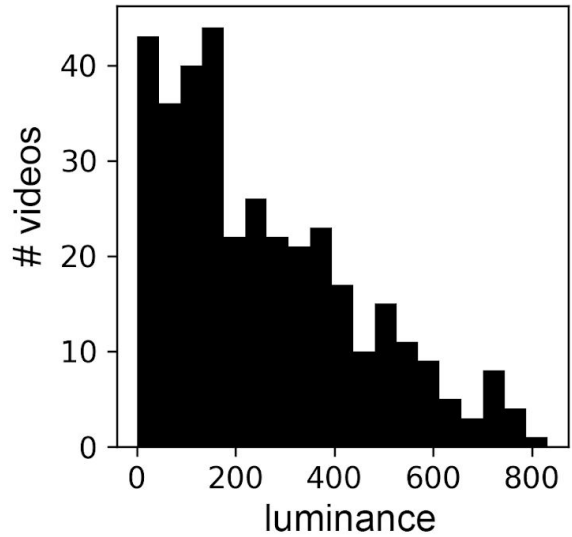

**(C)** Balance Error 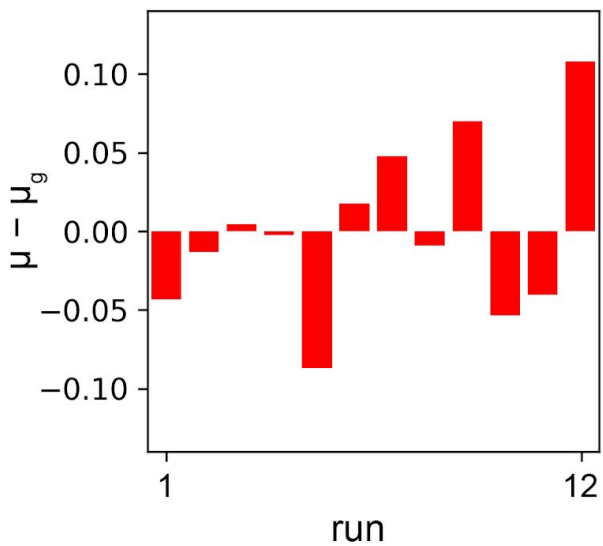

**(D)** MILP vs Randomization 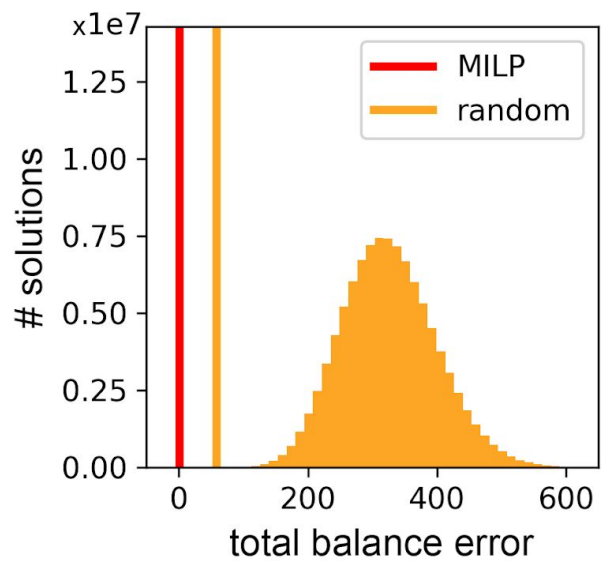

**Figure 2: MILP Finds Optimal Balanced Grouping**
**(A)** In this problem, a researcher needs to divide a dataset of 360 short video stimuli into 12 separate runs. The main challenge is dividing the videos in a way that makes the mean luminance of each run approximately equal. **(B)** This panel shows the distribution of mean luminance across the 360 videos. MILP is able to directly represent each of these 360 values to find the optimal grouping for this dataset. When MILP is used to perform the video grouping, the runs are well-balanced, as shown in **(C)**. Here the red bars represent the deviation between the mean luminance of each run and the global mean luminance. These values are very small compared to the overall range of luminances shown in (B), indicating that the runs are well-balanced. In this figure and the remaining figures of this section, red is used for information relating to an optimal solution found using MILP. **(D)** This panel shows a comparison of the grouping found by MILP to $10^8$ randomly generated groupings. The quality of each grouping is measured by total absolute error, which is the absolute sum of run deviations shown in (C). The best randomized solution (shown by the yellow vertical bar) has a total error that is more than 100 times larger than the error of the optimal solution found using MILP (shown by the red vertical bar).

## 4.2. Case Study 2: Stimulus-Task Pairing

Another common neuroimaging challenge is designing top-down attention experiments. In these studies, the subject's attention state is varied systematically to demonstrate the effect of attention on brain activity. If similar stimuli are presented during each attention condition, then attention is assumed to be the operative factor underlying any differences in brain activity.

Visual attention experiments are typically limited to a small number of attention conditions. Each additional condition requires more data to be collected and also adds complexity to the overall design. Each attention condition in an experiment might also have its own idiosyncratic requirements. For example, some visual tasks (e.g. object identification) can only be performed on certain types of images (e.g. images containing objects). These experiments must be carefully balanced to ensure that the only meaningful difference across each condition is the subject's attentional state.

In this case study, we demonstrate how MILP can address these challenges to create designs that have large numbers of attention conditions. Suppose a researcher is designing a visual attention experiment in a manner similar to (Clark et al. 1997), (O'Craven et al. 1999), or (Harel et al. 2014). In each of these studies, the subject's attention state is varied independently from a visual stimulus to demonstrate the effect of attention on brain activity. These studies used 2, 3, and 6 attention states, respectively. For this example, suppose the researcher would like to significantly increase the number of attention conditions in an attempt to build a richer and more complete model of how attention affects brain activity.

*Design Specifications*

The researcher would like the experiment to consist of many individual trials. During each trial, the subject will first be cued with a visual search target, such as an object category, a scene category, or a color. Then, an image will briefly flash. Finally, the subject will have a response period to indicate whether they detected the search target in the image. The researcher has allotted time for 2800 trials evenly split across 14 different search conditions, resulting in 200 trials per condition.

The researcher has a stimulus dataset of 700 images. Each trial will use one of these images. Each of the images has been labeled along each of the 14 different search dimensions with one of three values. A "0" indicates that the image definitely does not contain the search target, a "1" indicates that the image might contain the search target, and a "2" indicates that the image definitely contains the search target. For this example we will create a synthetic dataset of image labels generated from a multinomial distribution where feature values [0, 1, 2] have probabilities [0.5, 0.25, 0.25].

The main challenge that the researcher faces is deciding which images to use with each attention condition. There are three different types of balance that the researcher would like to impose. First, to reduce the effect of the subject memorizing the images, the researcher would like each image to appear in an equal number of trials throughout the experiment. This results in 4 trials per image (= 2800 trials / 700 images). Each image should also be paired with each attention condition no more than once. Second, the researcher would like to control for the effects of target detection (Guo et al. 2012; Çukur et al. 2013). To this end, the trials within each task should be evenly balanced across the 3 detection levels, meaning that ⅓ of trials should definitely contain the search target, ⅓ should ambiguously contain the search target, and ⅓ should definitely not contain the search target. Finally, to ensure that the stimulus feature distributions are similar across tasks, the researcher would like the feature distribution of each task to resemble the global feature distribution. More specifically, the mean value of each of the 14 features should be approximately equal across conditions.

*MILP Formulation*

This problem can be seen as a variant of the previous balanced grouping problem in **Section 4.1** where the groups are now task conditions rather than runs. The main differences are that: 1) each group has its own unique constraints, 2) stimuli are allowed to appear in more than 1 group, and 3) multiple features are being balanced across groups.

Use *i* to index images, *t* to index tasks, and *f* to index stimulus features. The main variable of interest is the pairing of stimuli with tasks. Introduce binary variables *X* to represent these pairings:

> VARIABLES
> $$X \in \mathbb{B}^{700 \times 14}$$
> $$x_{i,t} = 1 \rightarrow \text{ image } i \text{ is paired with task } t$$
> $$\phantom{x_{i,t}} = 0 \rightarrow \text{ image } i \text{ is not paired with task } t$$

It is simple to constrain each task to have the same number of trials (2800 / 14 = 200)

> CONSTRAINTS
> $$\sum_i x_{i,t} = 200 \qquad \forall \, t \in \{1, ..., 14\}$$

It is also simple to require that all images appear an equal number of times throughout the experiment. For 2800 trials and 700 images, each image should be used 2800 / 700 = 4 times.

> CONSTRAINTS
> $$\sum_t x_{i,t} = 4 \qquad \forall \, i \in \{1, ..., 700\}$$

For each task, there should be an equal number of trials where the search target is present, ambiguous, or absent. Thus each of these three feature levels should have 200 × ⅓ = 66.67 trials. Since this is not an integer, constraints can be constructed using the integral floor and ceiling of this number. Let $S_{t,v}$ be the set of stimulus indices that have feature value v for task t.

> CONSTRAINTS
> $$\sum_{i \in S_{t,v}} x_{i,t} \geq 66 \qquad \forall \, t \in \{1, ..., 14\} \text{ and } v \in \{1, 2, 3\}$$
> $$\sum_{i \in S_{t,v}} x_{i,t} \leq 67 \qquad \forall \, t \in \{1, ..., 14\} \text{ and } v \in \{1, 2, 3\}$$

The final part of the program is a cost function that promotes a similar stimulus feature distribution within each task. Use a matrix *L* to refer to the feature values of each stimulus, where $L_{i,t}$ is the value of feature *t* for image *i*. Also, let $\mu_f$ be the global mean value of feature *f* across all images. *L* and $\mu_f$ are constants predetermined by the image dataset. Use a cost function that minimizes the deviations between the global feature means $\mu_f$ and the feature means of each task $m_{f,t}$.

> COST FUNCTION
> $$m_{f,t} = \frac{1}{200} \sum_i x_{i,t} \, L_{i,f}$$
> $$\phantom{m_{f,t}} = \text{mean value of feature } f \text{ across trials of task } t$$

$$\min \sum_{\substack{f,t \\ f \neq t}} |m_{f,t} - \mu_f|$$

A solution to this program is shown in **Figure 3.**

**(A)** Stimulus Task Pairing

**(B)** Uses per Stimulus
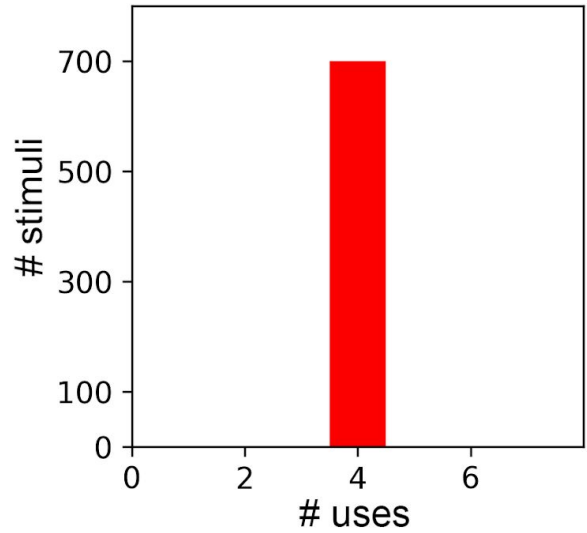

**(C)** Trials per Detection Level
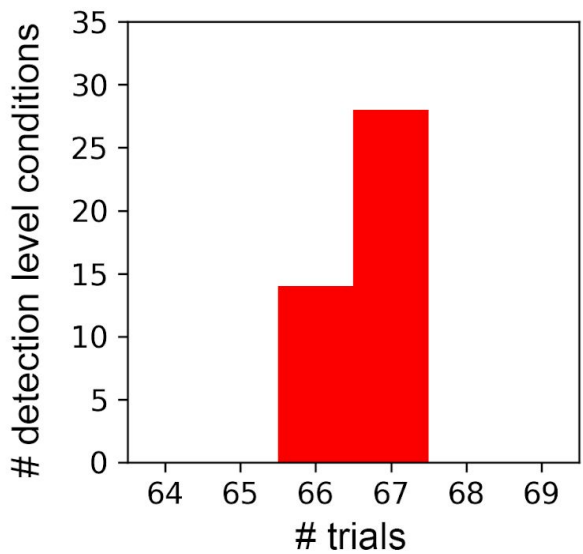

**(D)** Feature Level Errors
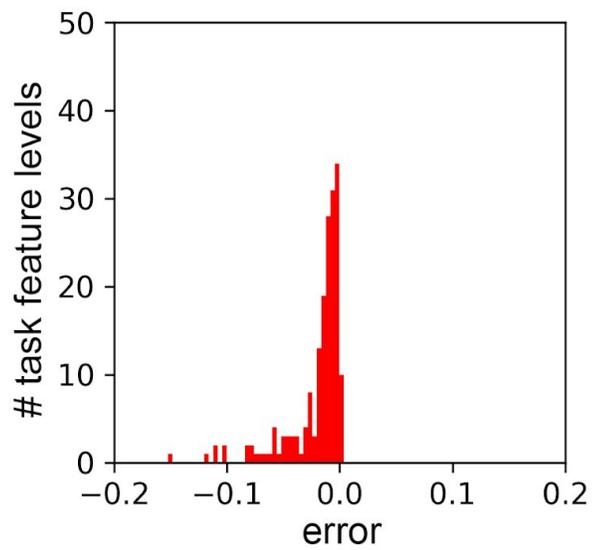

**Figure 3: MILP Finds Optimal Pairing Between Stimuli and Tasks**
**(A)** MILP is a natural fit for designing experiments that have specific rules that constrain how different stimuli and tasks can combine. Suppose a researcher is designing a visual search experiment with 14 search tasks of 200 trials each. Each trial will use an image from a dataset of 700 stimulus images. The researcher needs to decide which images to use during each task. They would like to impose three constraints on this pairing: 1) each image should be used the same number of times throughout the experiment, 2) the distribution of image features within each search condition should approximate the global feature distribution across all images, and 3) trials within each search condition should be evenly split across the 3 detection levels. Panels **(B)-(D)** describe properties of a solution to this problem found using MILP. **(B)** This solution uses each of the 700 stimuli the same number of times throughout the experiment. **(C)** This solution also evenly divides the trials of each search task across the 3 detection levels. Because the number of trials within each task must be an integer, an exact 3-way split is not possible. However, MILP is able to find the closest integral approximation. Each search task has 200 trials, and thus has 66 or 67 trials for each search condition. **(D)** The stimulus feature distribution within each task closely matches the global feature distribution. Each item in this histogram is the deviation between a mean feature value within a task and the global mean for that feature value. (The target feature within each task is omitted, since it is handled by a separate constraint, resulting in a set of 14 * 13 = 182 values). These errors are small compared to the feature values of 1, 2, and 3 that are used to label the stimuli.

## 4.3. Case Study 3: Structured Hierarchical Sampling

Another common neuroimaging design challenge is sampling stimuli from a highly structured space, such as the space of natural language. Such spaces are difficult to sample because samples must obey strict rules rather than being drawn from a simple probability distribution. Natural language stimuli cannot be generated by simply combining random words. For language to be intelligible, words must be jointly compatible in a meaningful way, obeying rules of grammar, syntax, and semantics. To sample from this type of space, one must be able to efficiently represent and navigate the rules of the space.

In this example, suppose a researcher would like to generate natural language stimuli in the form of questions about concrete nouns. These questions will be used in an experiment that asks a large number of questions about a large number of concrete nouns, in a manner similar to (Sudre et al. 2012). In this previous study, the authors asked 20 questions about each of 60 nouns in approximately 1 hour of scanning time. In this example we will use a sparse sampling strategy to increase the scope of this experiment to 5x the number of nouns and 6x the number of questions while only using 2x the trials. We will also allow the researcher to specify rules about which nouns are semantically compatible with each question.

Questions are formed by pairing a single concrete noun (e.g. a car, a phone, a flower) with a question template (e.g. How heavy is <X>? When was <X> invented? What color is <X>?). Concrete nouns are organized into a semantic hierarchy (**Figure 4B**). Unlike the (Sudre et al. 2012) study, each question template is only applicable to nouns from a particular portion of this hierarchy (**Figure 4C**). For example, questions related to object affordances might only be compatible with nouns from the "Inanimate Object" portion of the tree. Questions related to social behavior might only be compatible with the "People" portion of the tree. The researcher would like to independently model the effects of questions and concrete nouns. To this end they will place constraints on how often each is sampled, and the manner in which they are allowed to combine.

*Design Specifications*

In this example the researcher has allotted scanning time for 2400 trials. There are a total of 300 concrete nouns organized into a 15-group semantic hierarchy shown in **Figure 4B**. Each concrete noun exists in exactly one of the 10 leaf groups, and each leaf group contains exactly 30 concrete nouns. For each of the 15 example groups shown, the researcher has 8 question templates, for a total of 120 question templates. Each question template is applicable to some subset of the semantic hierarchy. Question templates related to a non-leaf group are compatible with any nouns for which that non-leaf group is an ancestor.

The main challenge that the researcher faces is deciding which concrete nouns to pair with each question template. Each noun should be paired with each question template no more than 1 time, and all pairings should respect the compatibility constraints of the semantic hierarchy. Each of the 120 question templates should be used the same number of times across the 2400 trials. Because each noun is compatible with a different number of question templates, it is not possible to exactly balance the number of times each concrete noun is used. However, noun usage should still be balanced as much as possible. Finally, for question templates that are compatible with multiple leaf groups in the semantic hierarchy, the researcher would like to balance the number of times each template is paired with each compatible leaf group.

*MILP Formulation*

Index question templates with $t$, concrete nouns with $n$, and noun leaf groups with $g$. Store information about template-group compatibility in a matrix $C$

$$C \in \mathbb{B}^{120 \times 10}$$
$$c_{t,g} = 1 \rightarrow \text{ question template } t \text{ compatible with group } g$$
$$\phantom{c_{t,g}} = 0 \rightarrow \text{ question template } t \text{ not compatible with group } g$$

Store information about noun-group membership in a matrix $M$

$$M \in \mathbb{B}^{10 \times 300}$$
$$m_{g,n} = 1 \rightarrow \text{ noun } n \text{ is in group } g$$
$$\phantom{m_{g,n}} = 0 \rightarrow \text{ noun } n \text{ is not in group } g$$

Matrices $C$ and $M$ are constants that are predetermined by the given semantic tree.

The main variables to be decided are pairings between question templates and concrete nouns. Introduce variables $P$ to represent these pairings

<u>VARIABLES</u>
$$P \in \mathbb{B}^{120 \times 300}$$
$$p_{t,n} = 1 \rightarrow \text{ question } t \text{ is paired with noun } n$$
$$\phantom{p_{t,n}} = 0 \rightarrow \text{ question } t \text{ is not paired with noun } n$$

Pairs $(t, n)$ for which $(CM)_{t,n}$ = 0 are invalid pairings. The corresponding $p_{t,n}$ variables can be set to 0 to reduce the size of the program. Each question template should be used 2400 / 120 = 20 times

<u>CONSTRAINT</u>
$$\sum_n p_{t,n} = 20 \quad \forall \, t \in \{1, ..., 20\}$$

The next step is to balance the number of times that each question template is paired with each of its compatible noun groups. The number of times that template $t$ is paired with group $g$ is given by an element of the matrix product $(PM^T)_{t,g}$. Let $v_t$ be the number of noun groups that are compatible with template $t$. Since each question template is to be used 20 times, each question template $t$ should be used (20 / $v_t$) times with each of its compatible noun groups. Since this quotient is not necessarily integral, $(PM^T)_{t,g}$ can be constrained to fall within its integral floor and ceiling.

<u>CONSTRAINT</u>
$$v_t = \sum_g c_{t,g}$$
$$\forall t \in \{1, ..., 120\} \text{ and } g \in \{1, ..., 20\} :$$

$$(PM^T)_{t,g} \geq floor\left(\frac{20}{v_t}\right)$$
$$\text{and}$$
$$(PM^T)_{t,g} \leq ceil\left(\frac{20}{v_t}\right)$$

Finally, a cost function can be created to balance the number of times each concrete noun is used. There are many ways this can be achieved. The number of times each noun is used is given by $a_n$. Suppose the researcher would like to avoid outliers, and thus wishes to minimize the range of noun occurrence counts. As explained in the **Section 3**, this range can be minimized as:

<u>COST FUNCTION</u>

$$a_n = \sum_t p_{t,n}$$
$$= \text{number of times noun } n \text{ is used}$$

$$\min \left\{ \max_n(a_n) - \min_n(a_n) \right\}$$

An optimal solution to this program is shown in **Figure 5**.

**(A)** Structured Sampling

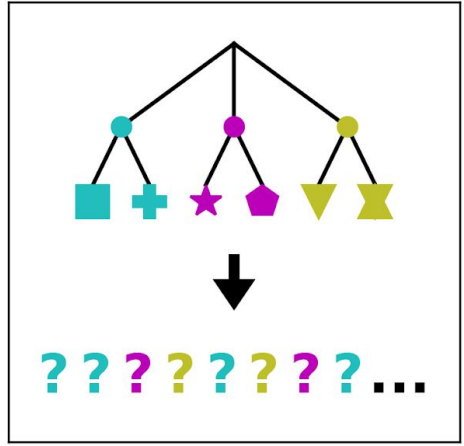

**(B)** Noun Semantic Tree

```
├── Entities
│   ├── Animals
│   ├── Corporations
│   └── People
│       ├── Famous People
│       └── Familiar People
├── Inanimate Objects
│   ├── Clothing Items
│   ├── Handheld Tools
│   └── Vehicles
└── Places
    ├── Buildings
    │   ├── Famous Buildings
    │   └── Generic Buildings
    └── Countries
```

**(C)** Question Templates

How old is <PERSON>?

Where is <PLACE>?

How expensive is <ANY>?

How big is <ANY>?

What does <CORPORATION> sell?

What job does <PERSON> have?

**Figure 4: Structured Spaces in Neuroimaging Experiments**
**(A)** In this example, a researcher is designing a question answering experiment. They would like to programmatically generate a set of questions using previously existing semantic structures, namely a semantic tree of concrete nouns and a set of question templates where each template is compatible with different branches of the tree. **(B)** The semantic tree used here has 15 total groups, with 10 leaf groups. Each leaf group contains 30 concrete nouns. **(C)** The concrete nouns will combine with question templates to produce questions for the experiment. Each question template is semantically compatible with nouns from a particular branch of the tree. Each of the 15 groups has 8 question templates. The main challenge of designing this experiment is ensuring that all of these features are balanced and combined in a controlled manner that respects the underlying semantic structure.

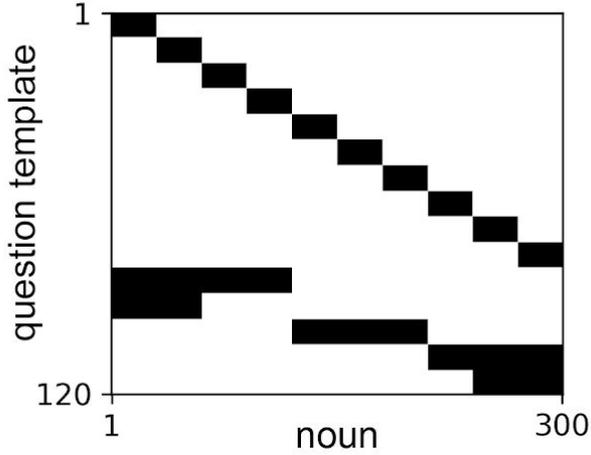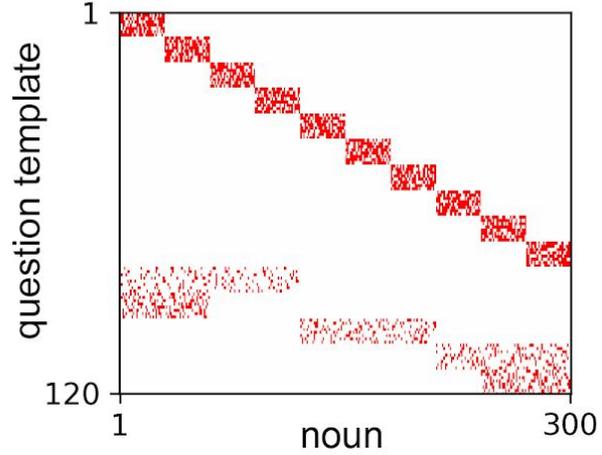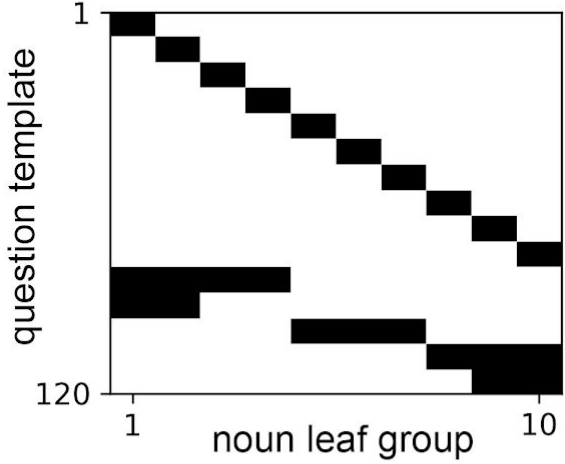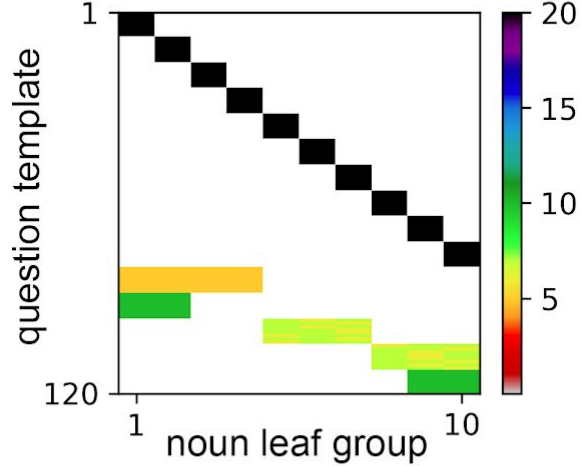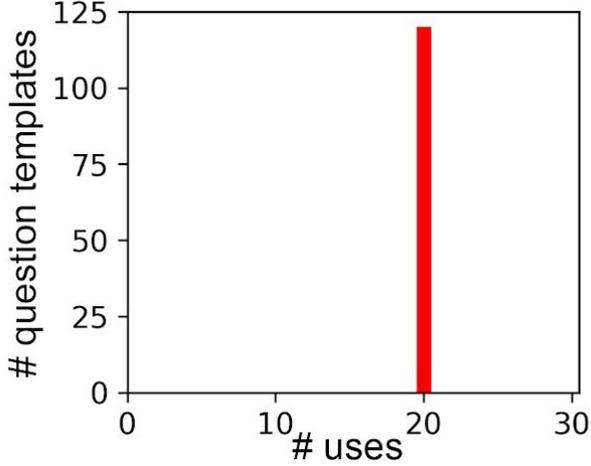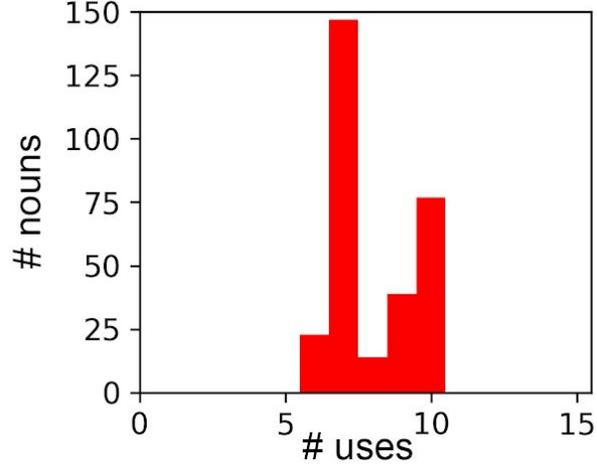

**Figure 5: MILP Finds Optimal Structured Sampling**
MILP is able to produce a set of questions that optimally samples the semantic structures of the previous figure. **(A)** The matrix *CM* describes the compatibility between pairs of question templates and concrete nouns. Here *CM* is shown with black values indicating compatibility and white indicating incompatibility. Each question template is compatible with between 30 and 120 concrete nouns. Each concrete noun is compatible with between 30 and 45 question templates. **(B)** The set of pairings found using MILP is shown in red. This is the matrix *P*. The pairings clearly form a subset of the possible pairings shown in (A). **(C)** The matrix *C* represents the compatibility between question templates and the leaf groups of the semantic tree of nouns. Since each of these leaf groups contains the same number of nouns, this matrix closely represents the matrix *CM* shown in (A). Each question template is compatible with somewhere between 1 and 4 leaf groups. **(D)** The matrix product $PM^T$ describes how question templates pair with noun leaf groups in the question set found using MILP. Each row of this matrix adds up to 20 and has a clear resemblance to the matrix *C* shown in (C). Importantly, MILP is able to find a solution that evenly splits the 20 trials of each question template among its compatible noun leaf groups. In this figure, this means that the non-zero entries of each row are values that are as equal as possible given the constraints of the problem. This helps prevent oversampling or undersampling some features combinations more than others. **(E)** The solution found using MILP uses each question template 20 times, as was specified in the constraints. **(F)** Because each noun is constrained to be compatible with a different number of question templates, it is not possible to find a solution that uses each noun the same number of times. However, MILP finds a solution that balances these frequencies as much as possible, given all the other constraints. The range of values [6, 10] is as small as possible given the constraints of the problem. Finally, we do not compare the solution found using MILP to an alternative randomized solution due to the complex constraints of this problem.

## 4.4. Case Study 4: Sequence Design for Navigation

Our final case study will demonstrate how MILP can address challenges associated with designing navigation experiments. Navigation is a rapidly advancing field of neuroscience and researchers are conducting studies that are increasingly rich and naturalistic (Spiers and Maguire 2006; Suthana et al. 2011; Nielson et al. 2015). As these studies grow more complex, so do the design constraints that must be integrated into their experimental designs. Here we show how MILP is a natural fit for representing and optimizing the structure of such experiments.

Suppose a researcher is designing a neuroimaging experiment where subjects must navigate a complex, naturalistic environment. Subjects will perform a "taxi driver" task where they are successively cued to drive to various locations on a map. Each time they reach a destination, a new destination cue will appear. Subjects will perform many of these trials throughout the course of the experiment.

An important aspect of this design is the sequence of cued destinations. This sequence will determine the particular distribution of navigational phenomena that the subject encounters throughout the experiment (Hartley et al. 2003; Xu et al. 2010). In the simplest case, this sequence could be generated randomly. However, this misses an opportunity to control the conditions measured by the experiment. Optimal selection of this sequence might require special consideration of the particular map being used and the hypotheses being tested.

*Design Specifications*

For this example, suppose a map has 25 possible destinations. The researcher would like to collect 80 trials per subject over the course of 40 minutes, resulting in a mean trial time of 30 seconds. To prevent memory effects related to the lengths of trials, the researcher would like path lengths of each trial to approximate an exponential distribution. To prevent memory effects related to repeatedly visiting locations, the researcher would like the number of times each location is visited to assume a geometric distribution.

*MILP formulation*

We will formulate this as a graph traversal problem where each location is a node and each route between locations is edge. This is similar to the classic traveling salesman problem, where the goal is to find a sequence that both visits every node once and minimizes total distance traveled. However, the goal here is instead to find a sequence whose edge length distribution maximally conforms to the target exponential distribution. Another difference is we would like to allow each destination to be visited more than once.

We will use notation $N_i$ to refer to node $i$, and $E_{i,j}$ to refer to the edge that connects $N_i$ to $N_j$. For simplicity, we will first formulate the problem where each node is visited at most once. Also for simplicity, we will randomly choose nodes $N_I$ and $N_F$ to be the initial and final nodes in the sequence. Introduce a binary variable to track the edges are used in the sequence

    VARIABLES
$$X \in \mathbb{B}^{25 \times 25}$$
$$X_{i,j} = 0 \text{ if edge } E_{i,j} \text{ is not used}$$
$$= 1 \text{ if edge } E_{i,j} \text{ is used}$$

Assume $X_{i,j} = 0$ for all $i$. We can constrain the number of edges in the sequence to equal the number of trials

    CONSTRAINTS

$$\sum_{i,j} X_{i,j} = \text{number of trials}$$

The number of times the subject enters and leaves each destination are given by the sums

$$\sum_i X_{i,j} = \text{ number of times entering } N_j$$

$$\sum_j X_{i,j} = \text{ number of times entering } N_i$$

For a well formed sequence, $N_I$ should be left once, $N_F$ should be entered once

CONSTRAINTS

$$\sum_i X_{i,F} = 1$$

$$\sum_j X_{F,j} = 0$$

$$\sum_j X_{I,j} = 1$$

$$\sum_i X_{i,I} = 0$$

Other nodes should be entered and left an equal number of times

CONSTRAINTS

$$\sum_i X_{i,h} = \sum_j X_{h,j} \qquad \forall\, h \notin \{I, F\}$$

To constrain trial length so that it is distributed exponentially, we will discretize exponential distribution into a 10 bin histogram, as shown in **Figure 6C**. Each bin represents a specific range of trial lengths $R_b$ and has a target number of trials $T_b$ that should fall in that range.

Let $L_{i,j}$ be the length of edge $E_{i,j}$. Let $A_b$ be the number of trials that fall within the range of bin $b$. Our cost function is then the deviation between the actual and target number of trials within each bin

COST FUNCTION

$$B_b = \{(i,j) | L_{i,j} \in R_b\}$$

$$A_b = \sum_{(i,j) \in B_b} X_{i,j}$$

$$\min \sum_b |T_b - A_b|$$

These constraints produce a sequence that is formed from $N_I$ to $N_F$. However, they also allow for the inclusion of "subtours", which are additional unconnected cyclic paths that exist alongside the main sequence. Since we want a single, acyclic sequence, we will utilize a common technique called "subtour elimination". Eliminating all subtours outright would require an intractably large number of constraints. Much more efficient is to iteratively solve a series of MILP programs, and successively add constraints until a solution free of subtours is found. Details of subtour elimination and iterative solving can be found in (Laporte and Nobert 1983). For each subtour detected in the intermediate solutions, we will add a constraint

CONSTRAINT

$$S = \text{ the set of edges in subtour}$$
$$\sum_{(i,j) \in S, i \neq j} X_{i,j} \leq |S| - 1$$

Finally, to allow each node to be visited multiple times, we will simply stack multiple copies of the original graph. Each node will be connected to all copies of all nodes other than itself. Any route through this augmented graph can be transformed into a route on the original graph by simply combining all copies of each node into a single node.

A solution to the problem as found by MILP is shown in **Figure 6.** Each location on the map is represented by a black dot (**Figure 6A**). The optimal sequence is shown by the red lines. Within this sequence, the target distribution and actual distribution of path lengths are shown in **Figure 6E.** The distribution of trial lengths exactly conforms to the target distribution. If it were not possible to exactly achieve this distribution, the MILP solver would find a sequence that conforms to this distribution as closely as possible. For comparison, we also generated $10^7$ random paths through the map, and we show the path length histogram for the random path that best conforms to the target distribution (**Figure 6F**). Unlike the solution found using MILP, the best randomized solution has substantial deviations from the target distribution. This shows that simple algorithms are not well-equipped to search this highly structured space, and that a more sophisticated framework such as MILP is needed to find optimal points.

Incorporating these diverse constraints would not be possible with traditional methods. Here MILP integrated structural constraints imposed by the map, time constraints imposed by data collection logistics, and behavioral constraints imposed by memory effects. Each design parameter can also be tweaked as needed in a flexible and explicit manner. For example, rather than simply controlling the trial length distribution, the researcher could control many other properties of the experiment by changing the nodes, edges, and other values associated with the graph structure.

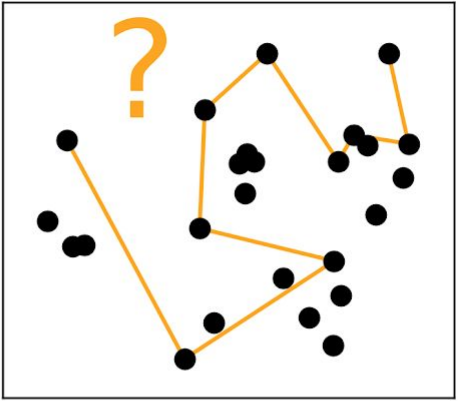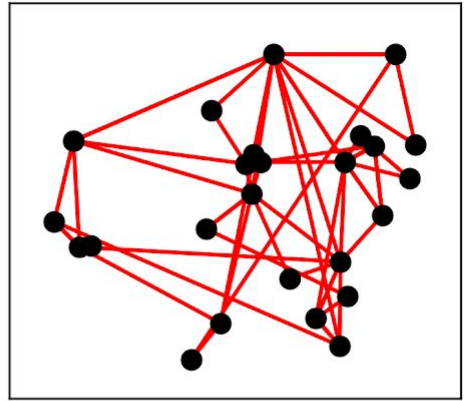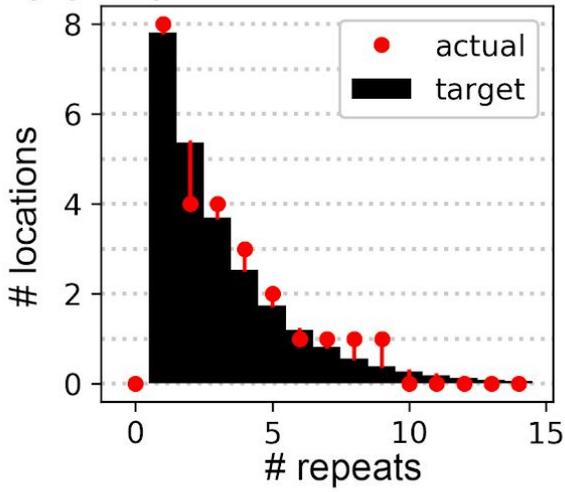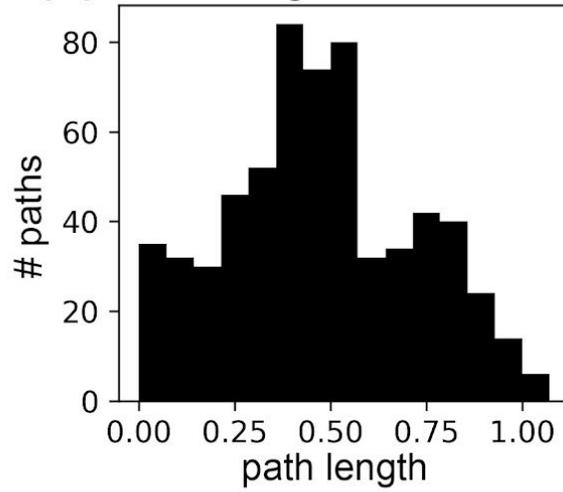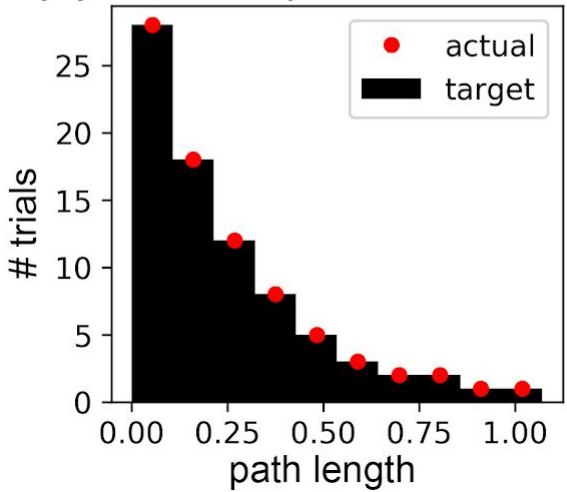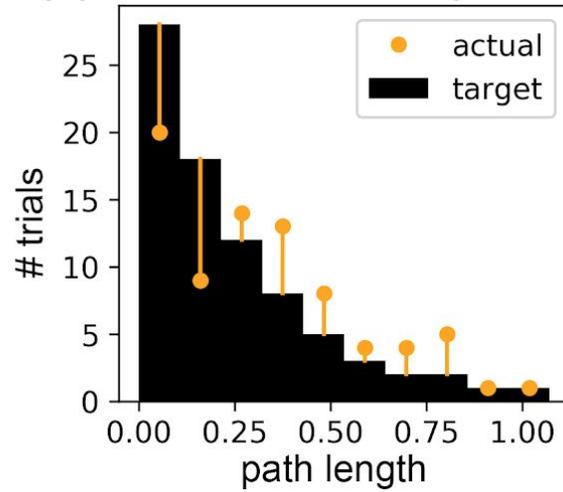

**Figure 6: MILP Finds Optimal Sequences for Navigation Experiment**
**(A)** Suppose a researcher is designing a navigation experiment where subjects must navigate along a particular sequence of locations on a map. The particular sequence used can affect which hypotheses the experiment can test, because that sequence determines the distribution of navigational features that the subject is exposed to. **(B)** MILP can optimize a sequence for testing a given hypothesis by combining neuroscientifically relevant constraints with the underlying spatial graph representation of the map. In this example, MILP is used to find a sequence of locations whose path lengths assume an exponential distribution. Also, the number of times that each location is visited is constrained to assume a geometric distribution. These particular distributions might be used if the researcher would like to prevent memory effects, but other distributions could be used based on the needs of the experiment. **(C)** MILP is used to determine how many times each location should be visited across the 80 trials. The distribution, shown by red dots, matches a target geometric distribution as closely as possible given that the number of samples is integral rather than continuous. **(D)** The map used in this example was randomly generated by placing 25 locations on a map using a uniform distribution. The resulting distribution of path lengths is shown. **(E)** The sequence of path lengths found using MILP exactly conforms to the target exponential distribution. If it were not possible to exactly achieve this distribution, the MILP solver would find a sequence that conforms to this distribution as closely as possible. **(F)** It is difficult to achieve the desired exponential distribution of path lengths through simple randomization. Of $10^7$ randomly generated sequences, this panel shows the one whose path length distribution most closely conforms to the target distribution. Large errors are apparent. The distribution bears a resemblance to the overall path length distribution shown in (D), suggesting that randomization is not well-suited for target path length distributions that are different from the original path length distribution.

# 5. Practical Considerations of MILP

## 5.1. MILP vs Other Experimental Design Tools

When discussing the challenges of designing complex experiments in **Section 2**, we mentioned some shortcomings of commonly used experimental design tools. Here we will discuss some of these tools in more detail and compare them to MILP.

### *Randomization*

A common approach for designing complex experiments is to randomly search the experimental design space for an optimal set of parameters. Randomized algorithms can be simple to implement and to use, especially for simple problems where all feasible solutions can be evaluated. However, randomization can become intractable in large experimental design spaces that have many diverse constraints. Randomization offers no general guarantee that the best randomized solution is optimal. Randomization also provides no way to tell how close the random solution is to an optimal one.

Even in our first case study, which was relatively simple, MILP finds a substantially better solution than randomization (see **Figure 2D**). It is possible that there might exist improved randomization techniques that better exploit the specific structure of these problems in order to search the parameter space more efficiently. However, MILP benefits from decades of algorithm development that allow many types of problems to be solved efficiently without much algorithmic tuning by end users. Compared to MILP, creating a custom randomization algorithm that is efficient and theoretically sound might require significant effort or expertise. In deciding whether to use MILP or randomization, one should consider whether each technique can fully represent all constraints of the problem, and whether each technique can provide a useful solution in a reasonable duration of time.

### *m sequences*

M-sequences are special sequences used to optimally order the items in an experiment (Sutter 1987; Benardete and Victor 1994; Buračas and Boynton 2002). M-sequences endow the ordering with many desirable properties related to efficiency and balance of conditions. However, m-sequences also impose undesirable rigid constraints on the problem and they are significantly less expressive than MILP. For example, m-sequences impose rigid constraints on how long the experiment can be, the number of conditions that can be used, and the ways that conditions must combine. These can be difficult to satisfy in a complex experiment where the parameters are somewhat predetermined by factors such as resource constraints or the scientific question of interest. The additional constraints imposed by m-sequences are especially problematic in high dimensional spaces. One area m-sequences excel over MILP is in developing sequences with optimal ordering properties. Using MILP to create a sequence of n items can require use of $n^2$ variables, with a unique binary variable for each possible position of each item. This makes MILP unsuitable for developing long sequences of items.

### *Combinatorial Designs*

Combinatorial designs provide prescriptions for sampling a space of experimental conditions. This might be accomplished by sampling all possible combinations of conditions (e.g. a fully factorial design), or by sampling a subset of combinations in order to optimize some specific property such as balance. Combinatorial designs are a broad class that includes latin squares (Friston et al. 1998; Hagberg et al. 2001), factorial designs (Friston et al. 1997; Hall et al. 2000; Gurd et al. 2002), and circulant arrays (Lin et al. 2017).

The benefit of combinatorial designs over MILP is that they can impart some forms of balance more efficiently, especially balance of interactions (see **Section 5.2** below). However, combinatorial design methods are significantly less expressive than MILP. They are not well-suited to constraining and optimizing multiple types of properties simultaneously, and they usually require that condition be treated identically in the design.

Combinatorial designs can also impose rigid limitations on the values allowed for their parameters (Vieira et al. 2011). Combinatorial designs would not be sufficient to design any of the four case studies of the previous section. It is important to note however that MILP can be used in conjunction with combinatorial designs to utilize the strengths of both methods. Examples include use of MILP with balanced incomplete block designs (Linden et al. 2004), supersaturated designs (Mandal and Koukouvinos 2014), fractional factorial designs (Sartono et al. 2015a), and orthogonal designs (Vieira et al. 2011; Vieira et al. 2013; Sartono et al. 2015b; Little et al. 2019).

*Optimal Designs*

Optimal Designs are a type of experimental design technique from statistics that are distinct from the notion of "optimal" that we use throughout this article. Optimal Designs are designs that most efficiently estimate the parameters of some predetermined statistical model, such as a regression model. Here, the structure of the model, its noise distribution, and the number of trials are predetermined by the researcher. The resulting Optimal Design will estimate the model parameters with minimal variance. Many neuroimaging studies have used Optimal Designs (Dale 1999; Wager and Nichols 2003; Kao et al. 2009; Kao et al. 2014).

Optimal Designs have the distinct advantage of being maximally efficient for estimating parameters for a given model. However, Optimal Designs address a different set of problems than MILP. Optimal Designs are optimized solely to efficiently estimate the parameters of a pre-determined statistical model. Optimal Designs also require an accurate noise model, which is not always available for complex neuroimaging experiments. In contrast, MILP is able to optimize a much wider array of quantities related to an experiment. Like combinatorial designs, Optimal Designs are less expressive than MILP and would not be sufficient to address the example case studies of the previous section of this article. There has, however, been some work to hybridize Optimal Design with constrained optimization (Sagnol and Harman 2015).

*Genetic algorithms*

Genetic algorithms are biologically-inspired heuristics that search a problem space using a population of solutions that exchange desirable traits. Genetic algorithms are often used as search algorithms for finding the Optimal Designs described in the previous paragraph but can also be used for other types of problems. Genetic algorithms are similar to MILP in explicitly framing experimental design as an optimization problem. Unlike MILP, there are many examples of genetic algorithms being used to design neuroimaging experiments (Wager and Nichols 2003; Kao et al. 2009). Compared to MILP, genetic algorithms can be computationally faster to use (Foster et al. 2014; Kuendee and Janjarassuk 2018). However, genetic algorithms often produce inferior optimization results and come with no guarantee of global optimality (Foster et al. 2014; Kuendee and Janjarassuk 2018). Additionally, genetic algorithms are usually less stable than MILP, producing different solutions each time the algorithm is run (Kuendee and Janjarassuk 2018).

## 5.2. Limitations of MILP

MILP has limitations that make it ill-suited for some classes of experimental design problems.

*Numerical Instability*

One of the failure cases of MILP is numerical instability. A MILP solver can encounter numerical instability when there are constraints that include quantities of vastly different orders of magnitude. For example, a constraint might require a very large quantity, such as the total number of milliseconds in an experiment, to have a precise relationship to a very small quantity, such as the number of milliseconds per frame in a movie. Performing floating point arithmetic on such quantities can lead to an accumulation of rounding errors that prevent successful completion of a solving algorithm. See (Klotz and Newman 2013a; Klotz and Newman 2013b) for examples of how precision and other forms of ill-conditioning can affect the stability of MILP.

*Scalability*

MILP is also inadequate when problems become too large. Here, problem size refers to the number of variables or constraints needed to represent the problem. Since MILP is NP-complete, the time needed to solve a mixed integer linear program can increase exponentially with the size of the problem (Cook 2010). However, there exists no general relationship between problem size and solution time, and there is no way to determine beforehand how long the solution process will take for a given problem. Instead this must be determined empirically. MILP is adequate for solving each of the case studies in this article, each of which required hundreds or thousands of variables and constraints.

*Expressivity*

MILP is also limited in that it cannot efficiently represent multiplications between non-binary variables. Such multiplications can be necessary for expressing second order metrics such as standard deviation or covariance. For example a researcher may wish to minimize the covariance between two experimental variables. Fortunately there exist many forms of constrained optimization other than MILP that can efficiently represent these nonlinear constraints. Some forms of optimization that can directly represent variable multiplications include Mixed Integer Quadratic Programming (Chaovalitwongse et al. 2005), Second Order Cone Programming (Alizadeh and Goldfarb 2003), and Semidefinite Programming (Vandenberghe et al. 2008). In this article we focus on MILP because it is capable of expressing a wide range of constraints that one might encounter in the design of neuroimaging experiments. However, researchers can use other types of constrained optimization if they encounter exotic experimental design constraints that are not efficiently expressed by MILP. It is important to note that this increased expressivity comes with a cost. Allowing for multiplications between variables can adversely affect the scalability and numerical instability of a problem. See (Burer and Saxena 2012) for an overview of the costs and benefits of various forms of constrained optimization and how they relate to one another.

## 6. Final Words

MILP provides a powerful framework for expressing design constraints that researchers encounter when designing complex or naturalistic fMRI experiments. The MILP framework is particularly adept at expressing many different types of constraints within a single formal representation. In this article we have demonstrated how MILP can be applied to four fMRI experimental design case studies that are not resolvable with standard experimental design techniques.